# Double-spiral magnetic structure of the Fe/Cr multilayer revealed by nuclear resonance scattering

M. A. Andreeva,[1,*] R. A. Baulin,[1] A. I. Chumakov,[2,3] R. Rüffer,[2] G. V. Smirnov,[3] Y. A. Babanov,[4] D. I. Devyaterikov,[4] M. A. Milyaev,[4] D. A. Ponomarev,[4] L. N. Romashev,[4] and V. V. Ustinov [4]

[1] *Faculty of Physics, M.V. Lomonosov Moscow State University, 119991 Moscow, Russia*
[2] *ESRF-The European Synchrotron, CS 40220, 38043 Grenoble Cedex 9, France*
[3] *National Research Centre "'Kurchatov Institute", Pl. Kurchatova 1, 123182 Moscow, Russia*
[4] *M.N. Mikheev Institute of Metal Physics RAS, 620990 Ekaterinburg, Russia*

We have studied the magnetization depth profiles in a $[^{57}\text{Fe}(d_{\text{Fe}})/\text{Cr}(d_{\text{Cr}})]_{30}$ multilayer with ultrathin Fe layers and nominal thickness of the chromium spacers $d_{\text{Cr}} \approx 2.0$ nm using nuclear resonance scattering of synchrotron radiation. The presence of a broad pure-magnetic half-order (½) Bragg reflection has been detected at zero external field. The joint fit of the reflectivity curves and Mössbauer spectra of reflectivity measured near the critical angle and at the "magnetic" peak reveals that the magnetic structure of the multilayer is formed by two spirals, one in the odd and another one in the even iron layers, with the opposite signs of rotation. The double-spiral structure starts from the surface with the almost antiferromagnetic alignment of the adjacent Fe layers. The rotation of the two spirals leads to nearly ferromagnetic alignment of the two magnetic subsystems at some depth, where the sudden turn of the magnetic vectors by ~180º (spin-flop) appears, and both spirals start to rotate in opposite directions. The observation of this unusual double-spiral magnetic structure suggests that the unique properties of giant magneto-resistance devices can be further tailored using ultrathin magnetic layers.

Magnetization depth profiles in superlattices consisting of alternating layers of ferromagnetic (FM) and nonmagnetic or antiferromagnetic (AF) materials attract nonvanishing interest since 1986 when it was discovered [1] that AF interlayer exchange coupling (IEC) between adjacent Fe layers across a Cr spacer leads to the giant magnetoresistance effect [2,3]. This discovery brought in 2005 the Nobel prize to A. Fert and P. Grunberg. There are two widely studied features of the magnetization arrangement between the FM layers. They are *(i)* the long-period and short-period oscillations of the AF IEC as a function of the thickness of a nonmagnetic spacer [4-7] (more Refs. in the review [8]), and *(ii)* the intriguing stair-case dependence of the hysteresis curves explained by the sequent layer-by-layer rotation of magnetization under the action of the external magnetic field $\mathbf{H}^{ext}$ [7, 9]. The latter observation destroys the simplest picture of the action of the increasing $\mathbf{H}^{ext}$ on the magnetization alignments in the Fe layers which includes, at first, the rotation to the perpendicular to the $\mathbf{H}^{ext}$ orientation of the AF coupled Fe layer magnetizations (spin-flop transition) [1, 3, 10], and afterwards the gradual uniform rotation of the two magnetic sub-systems to the direction of the external field. In some systems the 90º initial orientation of the two magnetic sub-systems have been discovered [11-14], supported by the bilinear-biquadratic formalism or specific proximity magnetism model ([15-16], and review [8]). However, the theoretical modeling and the more sophisticated experimental techniques, giving the depth resolved magnetization profiles, like polarized neutron reflectivity (PNR) or nuclear resonance reflectivity (NRR), present a more complicated picture of the depth resolved magnetization reorientations including the layer-by-layer twisting of the magnetization in each magnetic sub-systems [17-27].

For the existence of the AF IEC the spacer thickness should be well marched (e.g. in [Fe/Cr] case it should be equal to ~0.9÷1.2 nm [2,4]). However, the real value of the spacer thickness in a prepared multilayer differs very often from the nominal parameter due to the technological specificity. Besides the obtained interface quality and possible impurities essentially influence the IEC (see e.g. [28]). One could suggest that if the spacer thickness does not match neither AF nor FM IEC, the magnetization depth profiles of the [Fe/Cr] system could be as complicated as those observed for AF systems under an application of $\mathbf{H}^{ext}$. In other words, for the intermediate thickness of the spacer it is plausible to expect complicated magnetization profiles like fan structure, spirals with alternating sign of rotation [29], etc.

Here we present such result for [$^{57}$Fe/Cr]$_{30}$ multilayer with the thickness of Cr spacer intermediate between FM and AF IEC. The two magnetic spirals, relating to the odd and even Fe layers, with different signs of rotation have been revealed. The obtained magnetization profiles have not appeared in the numerous theoretical modeling of the AF superlattices [17, 23-25].

The studies were enabled due to the novel Synchrotron Mössbauer Source (SMS) [30, 31]. Different from a common radioactive source, the radiation coming from the SMS is the needle-like collimated beam with small (~ mm) size, which can be further focused to spot sizes of few micrometers [31]. It is important that the radiation from SMS is fully $\pi$-polarized. These properties make the SMS an ideal device for Mössbauer spectroscopy in the conventional energy scale in reflectivity geometry and supply us the rich information about the magnetization depth profiles in multilayers.

The series of $Al_2O_3/Cr(7\ nm)/[^{57}Fe(x\ nm)/Cr(y\ nm)]_{30}/Cr(1.2\ nm)$ samples with ultrathin $^{57}Fe$ layers (0.08 nm < x < 0.8 nm) and various Cr spacers (y=1.05 nm, 2.0 nm) was grown at the Katun'-C molecular beam epitaxy facility in the Institute of Metal Physics in Ekaterinburg. The measurements were performed at the Nuclear Resonance beamline [32] ID18 of the European Synchrotron Radiation Facility (ESRF). The storage ring was operated in multi-bunch mode with a nominal storage ring current of 200 mA. The energy bandwidth of radiation was first reduced down to 2.1 eV by the high-heat-load monochromator [33], adjusted to 14.4125 keV energy of the nuclear resonance transition in the $^{57}Fe$. Then x rays were collimated by the compound refractive lenses down to the angular divergence of few μrad. Final monochromatization down to the energy bandwidth of ~8 neV and the sweep through the energy range of about ±0.5 μeV was achieved using the SMS [30, 31]. Radiation from the SMS was focused down to the beam size of 8×10 μm² using the Kirkpatrick-Baez multilayer mirror system. The intensity of the x-ray beam on the sample site was ~$10^4$ photons/s. For varying temperature and $\mathbf{H}^{ext}$, the samples were mounted in the cassette holder and placed into the He-exchange gas superconducting cryo-magnetic system. The experimental data were analyzed with the program package REFSPC, developed specifically for these studies and uploaded to the ESRF scientific software web site [34].

Figure 1 shows the x-ray and NRR curves. X-ray reflectivity has been measured using the Renninger reflection option of the SMS [31], where SMS provides radiation in the energy bandwidth of ~10 meV. NRR curves have been obtained with the SMS energy bandwidth of ~8 neV which is continuously swept through the energy range of about ~0.5 μeV, i.e., the NRR curve represents the integral over the Mössbauer spectra of reflectivity at each incidence angle.

It was supposed that the sample having nominal 2 nm Cr spacers should have a FM ordering between $^{57}Fe$ layers. However, a small bump at the position corresponding to the half order (½) Bragg peak, which is absent in the X-ray reflectivity curve, appeared on the NRR curve for that sample. The half-order "magnetic" maximum indicates the presence of the magnetic structure with the period two times larger than the chemical period. Fig. 1b shows that applying $\mathbf{H}^{ext}$ first increases the "magnetic" peak, and then suppresses it. The fit of the X-ray reflectivity curve gives the period of the structure of 2.24 nm with 1.58 nm for the Cr layer thickness, which corresponds in reality to the intermediate value between FM and AF IEC.

Not only the "magnetic" maximum has the smeared shape, but a rather unusual distortion of the first order Bragg peak is seen on the NRR curve. Such a satellite-like distortion of the peaks evidences the more complicated than AF ordering of $^{57}$Fe layers. Indeed, the model calculations for various kinds of non-collinear magnetic ordering [35] show that, in order to have the existence of the "magnetic" maximum and simultaneously satellites near the first order Bragg peak, the magnetic structure of the multilayer should include partially spiral and partially AF alignments of the magnetic moments of iron layers.

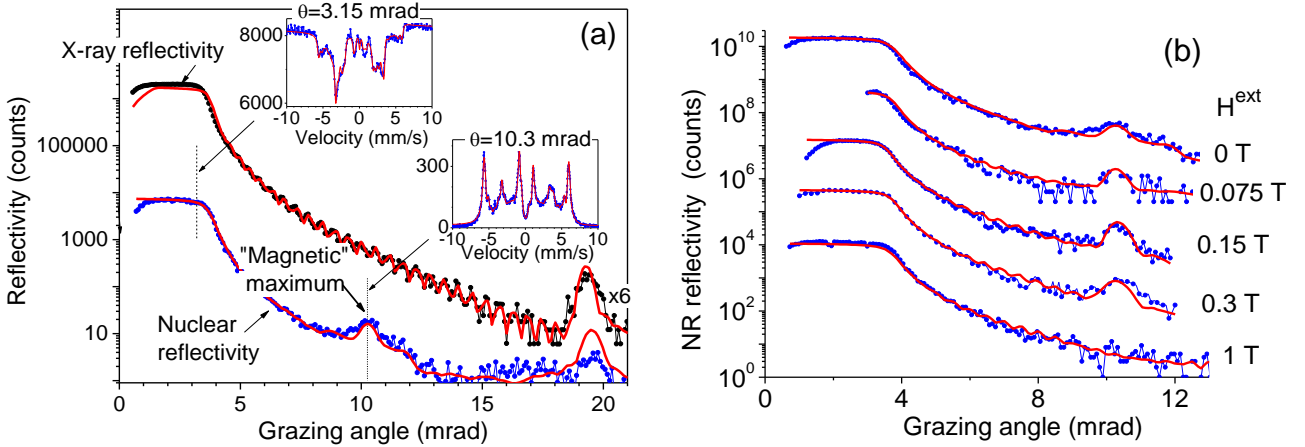

FIG. 1. (a) X-ray and NRR curves measured at 4 K and without $\mathbf{H}^{ext}$. Inserts: the Mössbauer spectra of reflectivity, measured near the critical angle of the total reflection and at the "magnetic" ½ peak. (b) The evolution of the "magnetic" ½ peak with $\mathbf{H}^{ext}$, applied perpendicular to the beam in the surface plane. Symbols are the experimental data, lines are the fits.

The measured Mössbauer spectra of reflectivity at the two angles of incidence (inserts in Fig. 1(a), and their change under $\mathbf{H}^{ext}$ application – Fig. 2) give the crucial information for the reconstruction of the magnetization profiles. For example, the comparison of Figs. 2c and 2d shows that the intensity of the 2$^{nd}$ and 5$^{th}$ lines of the spectra, corresponding to the $\Delta m = 0$ hyperfine nuclear transitions (located at ~± 3 mm/s), are noticeably higher at $\mathbf{H}^{ext}=0$ than at $\mathbf{H}^{ext}=0.3$ T. In the case of the symmetric relative to the x-ray beam direction alignment of the magnetic moments in the adjacent iron layers, these lines should be entirely suppressed in the Mössbauer spectra at the "magnetic" ½ peak [36]. Thus, their appearance in the measured spectra evidences the non-collinear and asymmetrical alignment of magnetization.

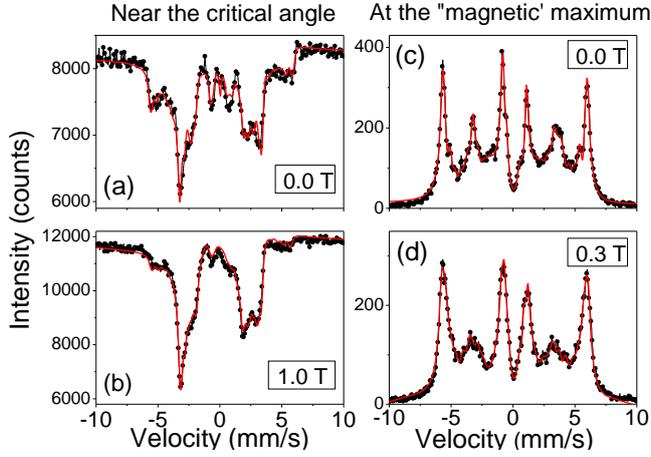

FIG. 2. Mössbauer spectra of reflectivity, measured near the critical angle (at ~3.15 mrad) and at the "magnetic" ½ peak (at ~10.3 mrad) for various $\mathbf{H}^{ext}$, applied perpendicular to the x-ray beam direction in the surface plane. The velocity scale is given relative to the Mössbauer spectrum of α-iron.

The joint fit of the x-ray, NRR curves and the Mössbauer spectra of reflectivity measured near the critical angle and at the "magnetic" ½ peak reveals the detailed pattern of the magnetic ordering of iron layers (Fig. 3).

In the absence of $\mathbf{H}^{ext}$ the magnetic structure forms two spirals, one for the odd and another one for the even iron layers, with the opposite signs of rotation, i.e., with opposite chirality (Fig. 3(a)). This double-spiral structure starts near the top of the surface from the almost AF alignment of the adjacent Fe layers. The rotation of two spirals in the opposite directions leads to the nearly FM alignment of the magnetic moments in the odd and even iron layers at some depth. Here the sudden turn of the magnetic moments in each magnetic sub-lattice by ~180° happens (the spin-flop effect). At larger depths, both spirals change the direction of rotation, still keeping the opposite sign of chirality.

Ramping $\mathbf{H}^{ext}$ to 0.075 T and 0.15 T leads to the noticeable increase of the "magnetic" ½ Bragg peak (Fig. 1a). Here the fit gives a simple picture of mainly the AF alignment of the magnetic vectors perpendicular to the direction of $\mathbf{H}^{ext}$. At stronger $\mathbf{H}^{ext}$ =0.3 T, the hyperfine magnetic field on the $^{57}$Fe nuclei begins to align antiparallel to $\mathbf{H}^{ext}$, but not jointly, this rotation starts from the top and bottom layers, where IEC is smaller (Fig. 3b). Note that the direction of the hyperfine magnetic field is opposite to the direction of iron magnetic moments. Therefore, the anti-parallel alignment of the hyperfine field relative to $\mathbf{H}^{ext}$ means the alignment of the iron magnetic moments along $\mathbf{H}^{ext}$. For yet stronger (but still relatively small) $\mathbf{H}^{ext} = 1.0$ T, the iron layer magnetizations set in pure FM alignment and the "magnetic" peak disappears (Fig. 1b).

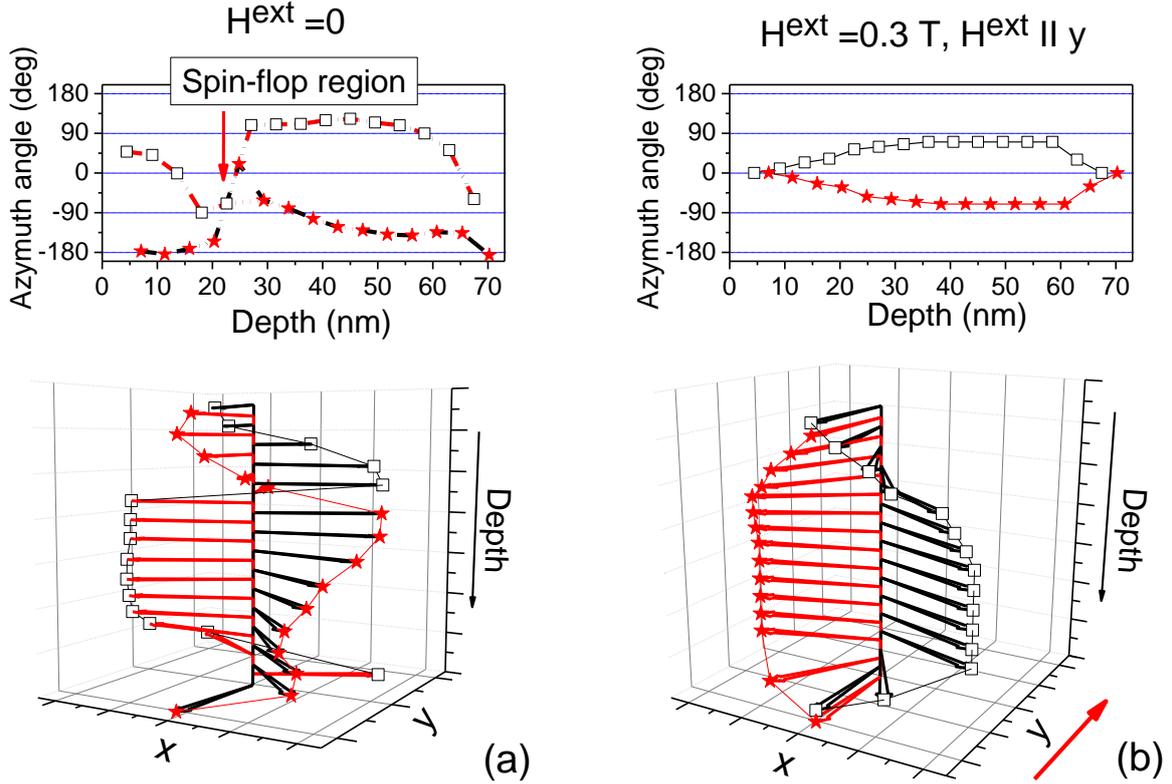

FIG. 3. The depth-profile of the orientation of the hyperfine magnetic field in iron layers without $\mathbf{H}^{ext}$ (a) and with $\mathbf{H}^{ext} = 0.3$ T, applied along the *y* axis (b). The asterisks mark even iron layers, whereas squares mark odd iron layers.

To the best of our knowledge, the double spiral magnetic structure with the opposite signs of rotations for two spirals and the alteration of the rotation direction with depth was not yet reported. Analyzing possible origin of this phenomenon, we note that the magnetization spin-flop most often appears for systems with an essentially large plane anisotropy and under an applied magnetic field [17, 23]. The studied here Fe/Cr system is characterized by the almost negligible plane anisotropy. This, most probably, rules out this reason from possible explanations.

On the other hand, the studied system is specific in the ultra-small thickness of the magnetic Fe layer and in the thickness of the Cr spacer, which is intermediate for FM and AF exchange coupling. Therefore, the observed unique non-collinear double spiral alignment of the AF lattices is possibly promoted by the attenuation of the IEC between the adjacent iron layers due to a rather thick spacing. This makes the interaction between the next neighbors (long range interaction) more important, and initiates the observed swirling magnetic ordering.

In summary, we studied the magnetic structure of the [Fe/Cr]$_{30}$ multilayer with the ultrafine thickness of Fe layers of ~0.66 nm and the thickness of the Cr spacer of 1.58 nm, which is an intermediate between the thicknesses optimal for FM or AF IEC. The obtained magnetization depth profile in iron layers shows up as a double-spiral stricture, where the spiral of the odd Fe layers is opposite in chirality to the spiral of the even layers. The result suggests that the unique properties of giant magneto-resistance devices can be further tailored using ultrathin magnetic layers.

From the methodological point of view, we demonstrate that the NRR method on the basis of the SMS provides us with the exceedingly wealthy information which gives the opportunity for an unambiguous determination of the magnetic structure with a single matching solution. It takes place because the measurements of the reflectivity curves – both with x-ray and nuclear resonance scattering – is accompanied by the measurements of the Mössbauer spectra of reflectivity for several incidence angles. The unique depth sensitivity of the Mössbauer reflectivity spectra, measured at different grazing angles provides the distinct advantage of nuclear resonance reflectivity in comparison to other techniques, e.g., to polarized neutron reflectivity.

The research was carried out within the state assignment of Russian FASO ("Spin" № 01201463330), was supported in part by the Russian Foundation for Basic Research (grants No. 15-02-01674-a, No. 15-02-01502-a, No. 16-02-00061-a and No. 17-02-00142-a) and by the Program of Ural Branch of Russian Academy of Sciences (grants No. 15-9-2-22).

*mandreeva1@yandex.ru